\documentclass[letterpaper,journal]{IEEEtran}
\usepackage{amsmath,amsfonts}
\usepackage{algorithmic}
\usepackage{algorithm}
\usepackage{array}
\usepackage{textcomp}
\usepackage{stfloats}
\usepackage{url}
\usepackage{verbatim}
\usepackage{graphicx}
\usepackage[caption=false,font=footnotesize]{subfig}
\usepackage{cite}
\usepackage{booktabs}
\usepackage{tabularx}
\usepackage{array}
\usepackage{ragged2e}
\usepackage{color}
\usepackage{eso-pic}

\newcommand{\IEEEAcceptedManuscriptNotice}{%
This article has been accepted for publication in \textit{IEEE Network}. This is the author's version which has not been fully edited and content may change prior to final publication. Citation information: DOI 10.1109/MNET.2026.3700795.}
\newcommand{\IEEECopyrightNotice}{%
\textcopyright~2026 IEEE. Personal use of this material is permitted. Permission from IEEE must be obtained for all other uses, in any current or future media, including reprinting/republishing this material for advertising or promotional purposes, creating new collective works, for resale or redistribution to servers or lists, or reuse of any copyrighted component of this work in other works.}

\AddToShipoutPictureFG{%
  \AtPageUpperLeft{%
    \put(0,-16){%
      \makebox[\paperwidth][c]{%
        \parbox{0.92\paperwidth}{\centering\fontsize{6.2}{7.2}\selectfont\IEEEAcceptedManuscriptNotice}%
      }%
    }%
  }%
  \AtPageLowerLeft{%
    \put(0,8){%
      \makebox[\paperwidth][c]{%
        \parbox{0.94\paperwidth}{\centering\fontsize{5.8}{6.7}\selectfont\IEEECopyrightNotice}%
      }%
    }%
  }%
}

\newcolumntype{Y}{>{\RaggedRight\arraybackslash}X}

\begin{document}

\title{Collaborative Orbital Edge Intelligence:  A Decentralized Paradigm for Energy-Efficient Computing in Space}

\author{Yuvraj Sahni, \textit{Member, IEEE}, Jiannong Cao, \textit{Fellow, IEEE}, \textit{and} Fu Xiao%
\thanks{Yuvraj Sahni is with College of Engineering and Computer Sciences, VinUniversity, Hanoi, Vietnam (Email: yuvraj.s@vinuni.edu.vn), Jiannong Cao is with the Department of Computing, The Hong Kong Polytechnic University, Hong Kong (Email: jiannong.cao@polyu.edu.hk), and Fu Xiao is with the Department of Building Environment and Energy Engineering, The Hong Kong Polytechnic University, Hong Kong (Email: linda.xiao@polyu.edu.hk).}}

\markboth{Accepted for publication in \textit{IEEE Network}. DOI: 10.1109/MNET.2026.3700795}%
{Sahni \MakeLowercase{\textit{et al.}}: Collaborative Orbital Edge Intelligence}


\maketitle

\begin{abstract}
In recent years, Low Earth Orbit (LEO) satellites have been increasingly deployed to enable connectivity in remote and disaster-prone areas. Researchers have proposed Orbital Edge Computing, which adds computational intelligence to LEO satellites to process data on orbit, providing edge intelligence close to space data sources. Existing work on Orbital Edge Computing typically assumes centralized control without collaboration among satellites from different providers, leading to limited connectivity, higher latency, and increased satellite battery depletion. They have not fully explored decentralized inter-satellite collaboration for energy-efficient intelligence under heterogeneous LEO constellations. This paper introduces a novel paradigm, Collaborative Orbital Edge Intelligence (COEI), that leverages decentralized collaboration among LEO satellites to enable energy-efficient on-orbit processing of space data. We describe the overall system architecture of COEI, including the issues related to networking, computing, and power management. COEI can help create a multi-party, multi-orbit megaconstellation of satellites that delivers better service quality by providing benefits, including global resilient connectivity, real-time intelligence, and energy-efficient services. To demonstrate the COEI benefits, we conduct a case study on decentralized energy-aware satellite task offloading to maximize the task success rate while minimizing the sum of the maximum battery depth-of-discharge across all satellites. Finally, we outline future directions for COEI that offer opportunities for further investigation. 
\end{abstract}


\section{Introduction}
\IEEEPARstart{T}{he} last few years have seen massive deployment of Low Earth Orbit (LEO) satellites by several companies, including SpaceX's Starlink, Amazon's Kuiper, and OneWeb by Eutelsat, that aim to provide global connectivity and high-speed broadband, especially for remote and disaster-prone areas. These private companies operate thousands of active LEO satellites and plan to launch more in the coming decade. There are many drivers of this growth, including policy shifts that enable private companies to deploy these satellites, technological innovation in reusable rockets, declining satellite costs, and rapid advances in hardware, communications, AI, and power management technologies \cite{bhardwaj2025ushering}. 


In addition to providing connectivity, these LEO satellites are used for several other applications, including climate sensing, agricultural monitoring, disaster response, maritime navigation, and missile tracking. These satellites generate large volumes of data that are transmitted to ground stations for analysis and decision-making, which can be costly, power-intensive, and subject to high latency due to limited bandwidth. Orbital Edge Computing (OEC) has emerged as an alternative paradigm that equips LEO satellites with computational capabilities via power-efficient edge computing hardware, enabling on-orbit data processing and machine learning inference in space \cite{denby2020orbital}. These satellites can process the generated data to provide low-latency application services and reduce bandwidth demands on downlinks. Researchers also envision using a constellation of LEO satellites equipped with computational capabilities to provide carbon-neutral data centers in space, rather than relying on energy-intensive terrestrial data centers \cite{aili2025development}. Cloud service providers, such as Google, also plan to equip satellites with processing units to create solar-powered AI data centers in space. 


Many studies in the literature have investigated various aspects of OEC, including constellation design, networking, task scheduling, power management, reliability, security, and their impact across different application domains.  Researchers have defined Satellite Computing as a broad paradigm \cite{guimaraes2025survey} in which satellites serve as compute-capable edge nodes. Collaborative satellite edge computing is another related paradigm that extends satellite computing by enabling tasks to be executed or offloaded across satellites, ground stations, and cloud servers. The work in \cite{jia2022collaborative} is an example in this direction, proposing a collaborative satellite-terrestrial network architecture and designing a central-edge multi-agent collaboration model. Phoenix \cite{liu2024orbit} is another popular solution that enables sunlight-aware task scheduling to minimize the maximum depth-of-discharge (DoD) of satellite batteries. More recently, another paradigm, called decentralized satellite networking \cite{oh2024call}, has been proposed, in which LEO satellites can form multi-party decentralized constellations to provide robust coverage. Some studies have also examined the problem of decentralized federated learning in distributed LEO satellite systems \cite{zhai2023fedleo}.

Existing solutions aim to address the resource and energy bottlenecks in LEO satellites by leveraging collaboration and task offloading. However, these works often rely on centralized or hierarchical control, which incurs high coordination overhead and is not feasible when satellites can belong to multiple providers. There are some works that leverage distributed multi-agent solutions, but they usually assume offloading scope across the satellite constellation and ground stations. However, implementing such offloading is inefficient in practice because satellites have limited access to up-to-date global network knowledge. Decisions made on current knowledge are ineffective in a dynamic network with high mobility and intermittent connectivity. Furthermore, research in this area is fragmented across three main directions: enabling local computing on satellites; collaboration through centralized/hierarchical offloading; and decentralized satellite networks with federated control for specific problems, such as networking and federated learning. Existing work has not fully explored the direction of decentralized inter-satellite collaboration for energy-efficient intelligence under heterogeneous LEO constellations.

In this paper, we propose a new paradigm, Collaborative Orbital Edge Intelligence (COEI), where LEO satellites across different orbital planes collaborate without a centralized controller to process on-orbit data and enable energy-efficient intelligence for applications. COEI with decentralized collaboration enables satellites from various providers, deployed across different orbital planes, to form a multi-party, multi-orbit megaconstellation of LEO satellites. COEI solution considers interrelated factors across multiple dimensions, including networking, computing, and power management. Table \ref{t:coei_comparison} shows the comparison of COEI with related paradigms. COEI enables satellites to make decentralized decisions based on local knowledge for energy-efficient connectivity and intelligence. It can provide several benefits, including global resilient connectivity, real-time intelligence, and energy-efficient services. It lays the foundation for achieving the vision of Ubiquitous AI, where anyone, anywhere, can have real-time access to intelligence.

\begin{table*}[t] \centering \caption{Comparison of Collaborative Orbital Edge Intelligence with related paradigms} \label{t:coei_comparison} \scriptsize \setlength{\tabcolsep}{4pt} \renewcommand{\arraystretch}{1.15} \begin{tabularx}{\textwidth}{p{2.5cm}YYYY} \toprule \textbf{Dimension} & \textbf{Orbital Edge Computing / Satellite computing} & \textbf{Collaborative satellite edge computing} & \textbf{Decentralized satellite networks} & \textbf{Collaborative Orbital Edge Intelligence} \\ 

\midrule 

Control Architecture & Centralized & Hybrid / Coordinated & Decentralized & Decentralized \\ 
\midrule 

Collaboration
& Limited 
& Tight 
& Federated
& Federated\\
\midrule

State Visibility
& Global
& Hybrid
& Local 
& Local \\
\midrule 

Provider Model
& Single-Operator
& Single / Coordinated Multi-Operator
& Multi-Operator
& Multi-Operator\\
\midrule 

Computation Role
& Primary - Onboard
& Primary - Cross-Tier
& Secondary 
& Primary - Cross-Tier \\
\midrule 

Compute-Network \mbox{Coupling}
& Compute-Oriented
& Joint Compute-Network
& Network-Oriented
& Joint Compute-Network\\
\midrule

Task Offloading Scope
& Local/Optional	
&Cross-tier	
&Limited/Optional	
& Neighbor + Ground \\
\midrule

Reliability Basis
& Infrastructure-Level
& Service-Level 
& Control-Level 
& Cross-Layer \\
\midrule 

Energy-Awareness
& Limited
& Joint
& Limited
& Joint \\
\midrule

Service Latency 
& Moderate
& Low 
& Moderate
& Low \\

\midrule 

Coordination Overhead
& Moderate
& High
& Moderate
& Moderate \\
\midrule

Energy-Efficiency 
& Moderate
& High
& Moderate
& High \\
\midrule 

Scalability
& Moderate
& Moderate
& High
& High \\

\bottomrule 
\end{tabularx} 
\end{table*}

To demonstrate the benefits of decentralized collaboration among satellites, we have conducted a case study on the decentralized energy-aware satellite task offloading (DESTO) problem in COEI, with the objective of maximizing the task success rate and minimizing the average maximum battery DoD across all satellites. Our proposed decentralized long short-term memory-based multi-agent proximal policy optimization (LSTM-based MAPPO) solution outperforms the baseline solutions in simulation experiments using real-world data from SpaceX's Starlink constellation. We also investigate the tradeoff between task success rate and battery DoD by varying the weights in the normalized cost function. Finally, we outline several future directions that offer opportunities for researchers to further investigate COEI. 

The main contributions of this paper are as follows:

\begin{itemize}
\item We propose a new COEI paradigm for decentralized collaboration among LEO satellites to enable energy-efficient on-orbit intelligence. We describe the system architecture and discuss the unique issues and benefits of COEI.
\item We conduct a case study on the DESTO problem in COEI. We propose and evaluate a decentralized LSTM-based MAPPO solution that leverages handcrafted local observation features, action masking, and objective-aligned reward design.
\item We outline several new future research directions and opportunities emerging from the COEI paradigm.
\end{itemize}


\section{Collaborative Orbital Edge Intelligence}

\begin{figure*}[!t]
\centering
\includegraphics[width=\textwidth]{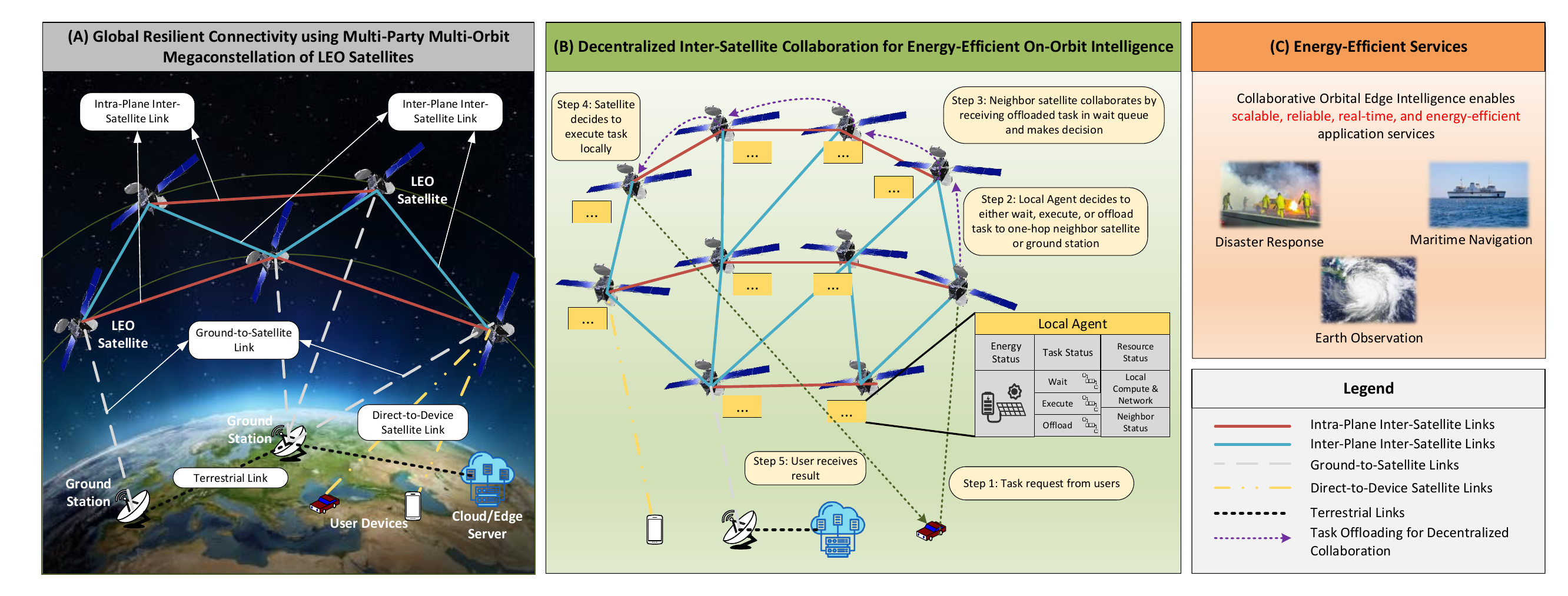}
\caption{(A) shows the overall architecture of Collaborative Orbital Edge Intelligence (COEI) with satellites in different orbits. The LEO satellites are connected in a mesh network via inter-satellite links, forming a multi-party, multi-orbit megaconstellation that provides resilient global connectivity. The satellites can also communicate with ground stations and user devices for collaboration. (B) shows decentralized inter-satellite collaboration in COEI by making local offloading decisions at each satellite. Each satellite can decide to wait, execute, or offload tasks to a one-hop neighbor or a connected ground station based on the energy, task, and resource statuses. (C) shows various applications, including disaster response, earth observation, and maritime navigation, that can be enabled by COEI. }
\label{f:coei_system}
\end{figure*}

Collaborative Orbital Edge Intelligence, shown in Fig \ref{f:coei_system}, is a new paradigm that leverages decentralized collaboration among LEO satellites to enable energy-efficient on-orbit intelligence. It addresses the limitations of traditional Orbital Edge Computing, which usually assumes a centralized controller to manage the constellation of satellites, by enabling decentralized control and collaboration among satellites from different providers/stakeholders. Different satellites can collaborate to share sensing, communication, or computation tasks, without collecting global private data. For example, a satellite can partition a complex AI model used to analyze Earth observation data into small tasks and offload them to neighboring satellites to reduce application latency and energy consumption. COEI enables the creation of a multi-party, multi-orbit megaconstellation of satellites, in which we can leverage satellites from different providers, deployed across various orbital planes, to provide maximum quality of service to application users. One key aspect of COEI is its emphasis on energy efficiency throughout the satellite's operations to maximize battery life. Later in Section \ref{s:casestudy}, we demonstrate this using a case study of the DESTO problem, where the application tasks are scheduled considering the tradeoff between the latency and battery DoD. 



\subsection{Main Issues}

\begin{figure}[!t]
\centering
\includegraphics[width=\columnwidth]{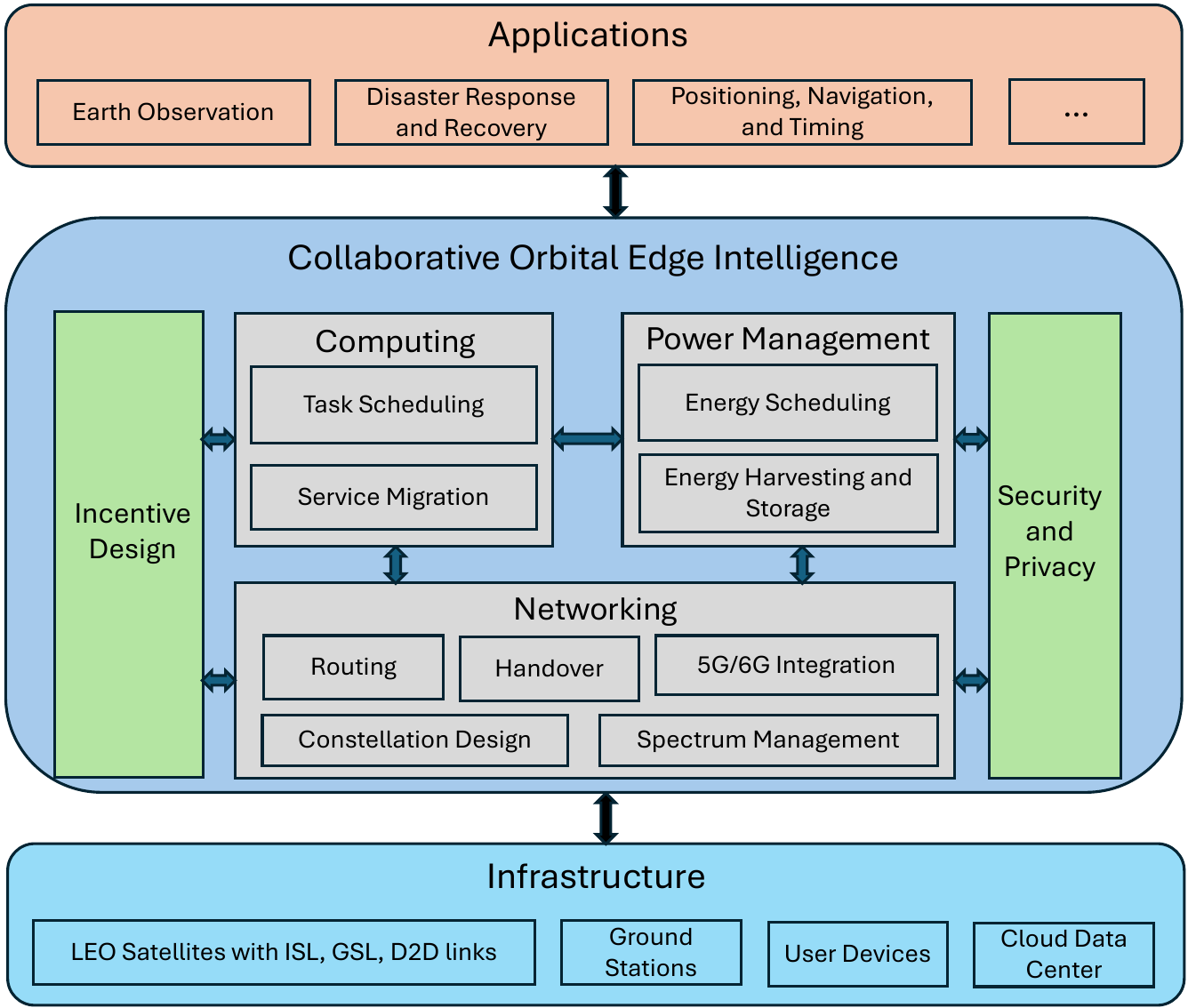}
\caption{Collaborative Orbital Edge Intelligence framework highlighting the main issues of networking, computing, and power management. Incentive design, security, and privacy are also essential to support decentralized collaboration in COEI.}
\label{f:coei_issues}
\end{figure}

In this section, we discuss the main issues in COEI across three interrelated dimensions: networking, computing, and power management, as shown in Fig. \ref{f:coei_issues}. 

\textbf{Networking}: Networking considers all aspects related to communication, including constellation design, spectrum management, routing, handover, and integration with 5G/6G networks. Constellation design involves determining several key factors, including the number of orbital planes, the number of satellites per orbital plane, orbital altitude, inclination, phasing, and right ascension of the ascending node spacing, based on the tradeoff among coverage, cost, and latency \cite{bhattacherjee2019network}. Spectrum management is another important issue as different satellites and terrestrial networks compete for limited frequency resources. Routing is also challenging because satellites are connected via dynamic ISLs that are vulnerable to factors such as weather, signal-to-noise ratio, power, and antenna gain. The high mobility of LEO satellites leads to frequent handovers, requiring new solutions in COEI, where satellites can be owned by different providers. Satellites in COEI can also be integrated with terrestrial 5G/6G networks, which raises the challenge of enabling seamless and secure routing as users connect to satellites and terrestrial networks from different providers. The key research issue in COEI is designing new mechanisms to provide global coverage and seamless, secure connectivity for users by integrating heterogeneous satellites from different providers into a megaconstellation. 


\textbf{Computing}: Computing in COEI involves all aspects of using and scheduling computational resources. In particular, task scheduling is a fundamental problem for satellites, in which each satellite decides how to partition the application tasks, when and where to execute them, and how to share satellite resources among multiple applications \cite{sahni2022distributed}. Another fundamental problem is service migration, in which applications must be dynamically migrated across satellites to provide a seamless user experience. Service migration can be particularly challenging for stateful applications that also require migrating their associated data. These problems are challenging because the satellites are resource-constrained, heterogeneous, highly dynamic, and prone to failures. COEI introduces additional challenges, as each satellite must make decentralized decisions without global knowledge of all satellites or their workloads. As different satellites may belong to multiple stakeholders, we must also consider incentivization, reliability, privacy, and security when making these scheduling decisions. The scheduling of computational resources across satellites also depends on other resources, including network, energy, and data. The key research issue in COEI is designing lightweight, decentralized algorithms and systems that jointly schedule computation and other resources across satellites to maximize application performance while considering system constraints.

\textbf{Power Management}: Power management in COEI encompasses energy harvesting, storage, and scheduling. Each satellite harvests solar energy to charge the battery and power its sensing, communication, and computing operations. The energy stored in the battery is used during the discharge cycle when the satellite is in the Earth's shadow region. However, the satellite's small size limits its energy harvesting and storage capabilities. Furthermore, the harsh space environment, including radiation, extreme temperatures, and other weather conditions, degrades the efficiency of energy harvesting and storage. Excessive energy use during the battery discharge cycle leads to a high DoD, reduces overall battery life, and increases costs. Therefore, it is essential to schedule satellite energy consumption to extend battery life \cite{li2024battery}. LEO satellites pose challenges for energy-consumption scheduling due to their highly dynamic environments and mobility, leading to frequent cycles of charging and discharging. COEI, with decentralized collaboration among satellites, introduces an additional challenge because satellites must account for the energy harvesting and storage schedules of other satellites. A key research issue in COEI is to develop predictive models for energy harvesting and storage that account for radiation and extreme temperatures. These predictive models should work in tandem with energy demand models to guide the scheduling and collaboration of sensing, communication, and computation tasks across the satellites.

\subsection{Potential Benefits and Applications}

The different satellites in COEI can collaborate to share the workload, adapt easily to dynamic network and workload demands, and provide scalable and reliable services, thereby improving the Quality of Service for application users. We have analyzed the benefits of COEI with respect to connectivity, intelligence, and energy efficiency.

\textbf{Global Resilient Connectivity}: Connectivity offered by existing LEO satellites is inconsistent and unreliable due to bent-type communication, frequent handovers, and environmental effects, including terrain, solar storms, rain, clouds, and temperature \cite{ma2023network}. To achieve global resilient coverage, more satellites are needed to cover different areas, and they should be able to collaborate. COEI, through decentralized collaboration among satellite providers, enables resource pooling to improve connectivity throughput and latency. Instead of relying on a single satellite provider to deploy a megaconstellation of thousands of satellites, COEI enables users to achieve global, resilient connectivity by leveraging satellite infrastructure from multiple providers to adapt to satellite and connectivity failures. COEI can help users leverage connectivity solutions from various satellite providers, including bent-pipe, inter-satellite, and direct-to-device communications, to achieve resilient connectivity in the face of environmental and other factors. For example, the satellites in the multi-party, multi-orbit megaconstellation can collaborate via ISL to efficiently route traffic, rather than relying on congested terrestrial links. COEI can help users access high-speed, resilient internet worldwide, especially in underserved regions. Several applications, especially disaster response and recovery, and emergency services, can benefit from COEI's global resilient connectivity.

\textbf{Real-time Intelligence}: The transmission of raw satellite data to ground stations suffers from intermittent connectivity and limited bandwidth, resulting in high latency. The work in \cite{denby2020orbital} showed that using edge computing on satellites can significantly improve latency, reliability, and scalability performance compared to a traditional bent-pipe communication architecture. Beyond the benefits of processing data locally across different satellites, COEI can achieve even lower latency by efficiently offloading sensing, communication, and computation tasks to satellites with sufficient resources throughout the megaconstellation. COEI can also support the deployment of complex AI models, such as large language models, by partitioning and scheduling them across satellites. Existing work has also shown the benefits of collaboration between satellites and ground stations in providing low-latency application services that can be integrated in COEI  \cite{liu2024orbit}. Several applications, including remote sensing, emergency response, real-time tracking, autonomous vehicles, and predictive maintenance for agriculture and industries, can benefit from the real-time intelligence enabled by the COEI paradigm. In the long term, COEI can enable Ubiquitous AI for users worldwide by allowing them to submit inference tasks to nearby satellites that provide real-time intelligence, rather than sending data to remote cloud data centers on Earth. 


\textbf{Energy-Efficient Services}: Each satellite is constrained by a power budget that limits its communication and computation capabilities. However, a megaconstellation of satellites developed using COEI can provide an energy-efficient infrastructure to support the demanding requirements of sensing, communication, and computation of emerging applications. Satellites with limited energy can collaborate to cover partial areas or offload tasks to satellites in orbits that receive sunlight.  Furthermore, satellites in COEI can offload tasks to terrestrial cloud data centers to maximize energy efficiency. The satellites can also leverage predictive energy-harvesting and storage models to decide whether to delay tasks without violating application and energy constraints. The study in \cite{aili2025development} showed that an orbital cloud data center built with a constellation of satellites, similar to a megaconstellation in COEI, can be more energy-efficient than a terrestrial data center, as satellites can harvest energy using solar power and use the vacuum as a cold sink. COEI can enable energy-efficient analysis of space-native data across a wide range of applications, including earth observation, disaster response, and positioning, navigation, and timing. Ground users can also offload data to satellites for energy-efficient machine learning inference, rather than relying on cloud data centers.

\subsection{Implementation Feasibility}
Decentralized collaboration in COEI requires integrating optical ISL, a capability that many providers, including SpaceX Starlink and the Galileo constellation, have demonstrated to be feasible. However, achieving accurate pointing, acquisition, and tracking with optical ISLs is challenging due to the harsh space environment and high mobility. Furthermore, a uniform standard is required for communication among different providers. Space Development Agency has established an optical ISL interoperability standard; however, it has not been adopted as a universal commercial standard. Besides communication, trust and incentivization are other concerns for collaboration among providers in COEI. Blockchain and other key enabling technologies for decentralized physical infrastructure networks \cite{lin2024decentralized} can be adopted in COEI to foster trust and collaboration among providers. Another work in \cite{wang2025space} discusses the implementation feasibility and challenges of designing space computing constellations in more detail, which also apply to COEI. 

It is also currently feasible to integrate LEO satellites with terrestrial 5G/6G networks. 3GPP identifies LEO satellites as a type of non-terrestrial network and provides specifications for integrating them with 5G. In particular, 3GPP has defined a regenerative payload mode in which the satellites have base-station functions on board and can use ISL for communication. 3GPP also considers service continuity for users being served by satellites from different providers. Release 17 by 3GPP also specifies meshed connectivity between satellites interconnected with ISLs. However, further efforts are required to define a clear framework for decentralized collaboration among satellites from multiple providers. The work in \cite{darwish2022leo} provided a detailed review of LEO satellites from a standardization perspective.



\section{Case Study} \label{s:casestudy} 

In this section, we present a case study on the DESTO problem, which considers decentralized energy-aware iterative task offloading across a mesh network of satellites in COEI.

\subsection{DESTO Problem Definition}

The DESTO problem can be defined as follows: Assume the time is slotted, and we are given a constellation of satellites represented as a connected graph, \textit{G = (V, E)}, where \textit{V} is the set of LEO satellites, and \textit{E} is the set of inter-satellite links connecting them. The satellites are heterogeneous, and each has maximum computing power, bandwidth, and battery capacity. Tasks arrive probabilistically at satellites according to a Bernoulli distribution at the start of each time slot. Each task is characterized by a computational load, a data size, and a deadline. We assume that task completion times can exceed a single time slot, implying that more than one task may be waiting to be executed in the queue at each satellite. The tasks can be scheduled at each satellite according to the predetermined priority order. In this paper, we order the tasks based on the Earliest Deadline First scheduling scheme. If tasks have the same deadline, we break ties by checking whether each task is at risk of expiry in the current time slot, and then use Shortest Job First to determine priorities. 

The objective of the DESTO problem is to maximize the task success rate and minimize the average maximum battery DoD across all satellites, subject to constraints on computing, bandwidth, battery capacity, and task deadlines.The DESTO problem involves making a decentralized decision at each satellite about whether to keep the task waiting, execute it locally, or offload it to a one-hop neighboring satellite. If the task is being offloaded, the satellite also decides where to offload it.

\subsection{LSTM-based MAPPO Algorithm}

The DESTO problem is modeled as a decentralized Partially-Observable Markov Decision Process (Dec-POMDP), in which each satellite agent uses its local observations to select an action and receives a globally shared reward. The main challenge in solving DESTO is that each agent can only observe local one-hop information, but the task can be iteratively offloaded to multiple hops. Furthermore, it is challenging to learn decentralized policies for a large number of agents in a complex COEI environment with heterogeneous satellites with varying computing and bandwidth capacities, dynamic and heterogeneous task workloads, and varying battery capacity and dynamics. 

We have proposed an LSTM-based MAPPO algorithm to solve the Dec-POMDP.  We have designed handcrafted observation features that capture the complex COEI environment, leveraged action masks that remove infeasible actions and improve exploration, and designed an objective-aligned reward function to enable effective policy learning. The local observation at each agent includes the satellite index and resource status, selected task information, one-hop neighbor features, and delayed summaries from one-hop neighbors. Action masking is leveraged to rule out all actions when no task is available and to set the action to keep the task waiting when the battery constraint is violated. The agents receive a normalized positive reward at every step for successful task completion and a negative reward for the average running maximum DoD. 

The proposed LSTM-based MAPPO uses shared actor and shared critic networks, making it easier to learn a policy in a network with a large number of agents. Furthermore, both the actor and critic networks use an LSTM layer that can learn from temporal patterns.


\subsection{Evaluation}

We have conducted a comprehensive evaluation of LSTM-based MAPPO through simulation experiments based on orbital mechanics, using real-world data \cite{liu2024orbit}. We have done an evaluation for two different constellations. The first one is a simulated constellation of 10 orbital planes, each with 10 satellites, and the second one is a realistic constellation based on Starlink \cite{liu2024orbit}, with 72 orbital planes, each with 22 satellites. The satellites are assumed to be connected using a +Grid topology \cite{bhattacherjee2019network}. The simulation has been conducted in a heterogeneous COEI environment with parameter values as shown in Table \ref{t:parameters}. 


The comparison has been done with four baselines: LOCAL, which always chooses the local execution action; GREEDY, which selects either wait, execute, or offload based on a score designed using a weighted task success rate proxy and the increase in the maximum value of DoD; PHOENIX-based solution, which is a decentralized adaptation of strategy in \cite{liu2024orbit}, where the offloading decision is restricted to one hop neighbor; and MAPPO, which is same as LSTM-based MAPPO except the LSTM layer. 

\begin{figure*}[!t]
\centering

\subfloat[Training Reward Curve of LSTM-based MAPPO]{
    \includegraphics[width=0.3\textwidth]{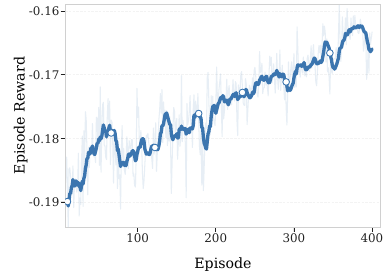}
    \label{f:training}
}
\hfill
\subfloat[Overall Reward during Evaluation]{
    \includegraphics[width=0.3\textwidth]{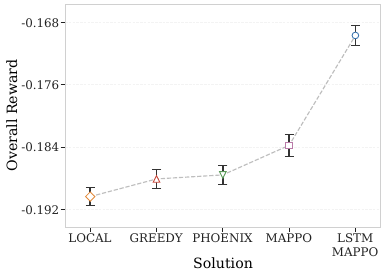}
    \label{f:evaluation}
}
\hfill
\subfloat[Task Success Rate vs Average Max DoD during Evaluation]{
    \includegraphics[width=0.3\textwidth]{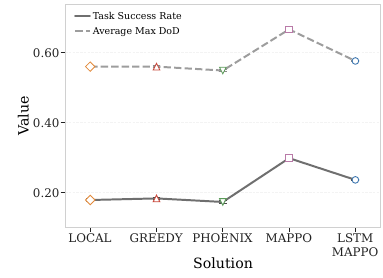}
    \label{f:tradeoffsuccess}
}

\caption{Performance comparison for the Starlink-based \textit{72$\times$22} constellation for the case of equal weights given to task success rate and maximum DoD in the normalized objective function}
\label{f:starlink}
\end{figure*}

Fig \ref{f:starlink} shows the performance evaluation for the Starlink-based simulation. The upward trend in the training reward curve (Fig. \ref{f:training}) indicates that our LSTM-based MAPPO policy learns effectively. The evaluation in Fig \ref{f:evaluation} shows that LSTM-based MAPPO achieves approximately 10\% higher overall reward than the LOCAL, GREEDY, and PHOENIX baselines. Fig \ref{f:tradeoffsuccess} shows that LSTM-based MAPPO achieves a much higher task success rate without compromising maximum DoD performance. Even compared to MAPPO, LSTM-based MAPPO achieves a 7.6\% higher overall reward.

\begin{figure*}[!t]
\centering

\subfloat[Task Success Rate]{
    \includegraphics[width=0.3\textwidth]{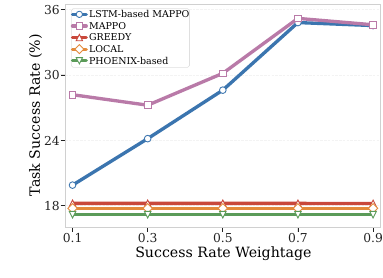}
    \label{f:task_success}
}
\hfill
\subfloat[Average Max DoD.]{
    \includegraphics[width=0.3\textwidth]{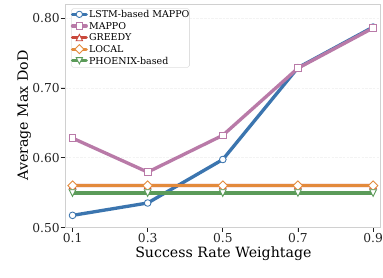}
    \label{f:dod}
}
\hfill
\subfloat[Overall Reward]{
    \includegraphics[width=0.3\textwidth]{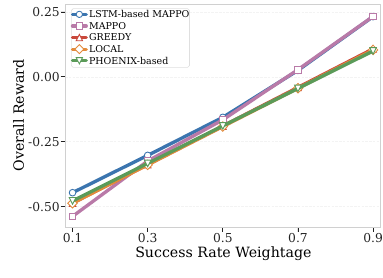}
    \label{f:reward}
}

\caption{Performance trade-off between task success rate and average maximum DoD for a \textit{10$\times$10} constellation under different objective weights}
\label{f:tradeoff}
\end{figure*}

Fig. \ref{f:tradeoff} shows that LSTM-based MAPPO achieves an efficient trade-off between task success rate and average maximum DoD, leading to a better overall reward than the baselines. Our proposed solution achieves, on average, reward improvement of over 60\% compared to LOCAL, GREEDY, and PHOENIX-based baselines. Both LSTM-based MAPPO and MAPPO can learn a policy that achieves a higher task success rate without significantly increasing the average maximum DoD. Even compared to MAPPO, LSTM-based MAPPO achieves, on average, 4\% and up to 17\% better reward performance.

 \begin{table}[t]
\caption{Parameter settings used for evaluation}
\label{t:parameters}
\centering
\scriptsize
\setlength{\tabcolsep}{3pt}
\renewcommand{\arraystretch}{1.1}
\begin{tabularx}{\columnwidth}{|p{0.42\columnwidth}|X|}
\hline
\textbf{Parameter} & \textbf{Value} \\
\hline
Satellite Computing Capacity & [100, 500] MFLOPs/s \\
\hline
Intra-Orbit Bandwidth Capacity& [900. 1900] MB/second \\
\hline
Inter-Orbit Bandwidth Capacity& [220.0, 650.0] MB/second\\
\hline
Battery Capacity & [60, 100] Wh \\
\hline
Initial Battery Percentage & [35, 70]  \\
\hline
Battery Charge Rate & [32, 48] W \\
\hline
Battery Idle Drain Rate & [3.2, 4.8] W  \\
\hline
Computing Power Consumption & [48, 72] W  \\
\hline
Transmission Power Consumption & [8, 12] W \\
\hline
Task Arrival Probability & 0.35 \\
\hline
Task Computing Load & [1000.0, 2600.0] MFLOPs \\
\hline
Task Data Size & [40, 110] MB \\
\hline
Task Deadline & [18, 32] seconds \\
\hline
Total Time Slots for Simulation & 3000 \\
\hline
Time Duration Per Slot & 2 seconds \\
\hline
\end{tabularx}
\end{table}
 
\section{Conclusion and Future Directions}

This paper proposes Collaborative Orbital Edge Intelligence, a decentralized computing paradigm for energy-efficient computing in space. COEI can leverage satellite infrastructure from multiple providers to deliver better, more resilient global connectivity, real-time intelligence, and energy-efficient services to users. However, COEI requires addressing several interrelated issues across networking, computing, and power management. Our preliminary case study on DESTO demonstrates the potential of leveraging decentralized collaboration to improve task success rate while reducing the maximum DoD of satellite batteries. COEI also presents many opportunities for researchers to address the following novel and challenging issues:

\textbf{Energy-Aware Joint Network and Computation Scheduling}:  COEI requires designing new decentralized algorithms to jointly schedule networking and computing resources considering the dynamic energy harvesting and storage schedule of satellites. The joint problem is challenging because several decision variables related to spectrum management, routing, and task scheduling are interdependent. For example, offloading tasks across different satellites without accounting for data transmission can lead to network congestion and degrade performance. Additionally, the high dynamics of satellites, uneven resource demand and utilization, and their susceptibility to failures make the problem more difficult, as scheduling solutions must predict and adapt to future network conditions. Moreover, the impact of energy harvesting and storage of heterogeneous satellites adds another dimension to the problem. The satellites must make these complex, interdependent decisions without global knowledge, thereby providing an opportunity for researchers to propose new approaches. 

\textbf{Reliable Stateful Migration of Applications}: As satellites orbit Earth, it becomes necessary to migrate applications between satellites or between satellites and ground stations to provide seamless services to users. Stateful applications pose additional challenges, as application data must also be migrated without disrupting service. Ensuring reliable migration of stateful applications between satellites is challenging due to their high mobility, which imposes stringent latency requirements. Additionally, satellites and their ISLs are prone to failures, requiring adaptive solutions. Novel live-stateful migration solutions for COEI are required that account for highly dynamic, resource-constrained satellites. 

\textbf{Incentive Design for Decentralized Collaboration}: COEI involves decentralized collaboration among satellites from different providers, requiring the design of appropriate incentive mechanisms to share the resources. COEI requires novel solutions that leverage game theory and blockchain-based approaches to encourage active collaboration. Furthermore, the incentive design should account for the sharing of diverse resources, including data, sensing, communication, computation, AI models, AI agents, and application services. In addition to satellites and their providers, the incentive design should consider other stakeholders, including service users and terrestrial network infrastructure providers. For example, how to price satellite services for users and 5G/6G network providers. A comprehensive solution is required to incentivize stakeholders to participate in the development and operation of COEI. The incentivization solution also requires a thorough economic analysis of the deployment and operation of LEO satellites in COEI to provide seamless services to users.


\textbf{Lightweight Platform for LEO Satellites}: Another important challenge for COEI is designing a platform, including an operating system, that integrates different algorithms and mechanisms. The platform should be lightweight and support integration with multiple satellite systems, including the resource virtualization and management system, the power management system, the navigation system, and the application interface. The platform should be modular and flexible, allowing updates to individual systems without affecting the overall platform. It should also support integration with emerging technologies, including software-defined networking, virtualized network functions, and AI-based autonomous systems.  A related challenge is also designing simulators and real testbeds to demonstrate the efficacy of the lightweight platform and the proposed mechanisms. 

\textbf{Security and Privacy}: Security and Privacy are essential for COEI to enable collaboration among satellites that are controlled by different providers. Existing solutions for decentralized collaboration among multiple stakeholders cannot be directly adapted to COEI, as satellites have frequent handovers and require real-time solutions. In particular, new decentralized identity solutions are needed to authenticate data, computing, and other services shared across satellites without relying on a centralized certificate authority. Homomorphic encryption is another approach that enables computation on encrypted data without compromising privacy. To ensure transparency and immutability among satellites, blockchain-based solutions can be adapted. However, these solutions should have low overhead to run efficiently on resource-constrained satellites. Quantum-resilient security and privacy solutions also need to be investigated for COEI, considering the rapid advances in quantum cryptography. 


\section*{Acknowledgments}
The authors acknowledge the use of Grammarly to improve the syntax, grammar, and academic tone of this manuscript, and of OpenAI ChatGPT to generate and debug Python scripts for performance evaluation in Section III. All content was independently reviewed, revised, and finalized by the authors, who bear full responsibility for the integrity and accuracy of the work.

\bibliographystyle{IEEEtran}
\bibliography{refer}

\section{Biographies}

\begin{IEEEbiographynophoto}{Yuvraj Sahni} (Member, IEEE) (yuvraj.s@vinuni.edu.vn) received the B.E. (Hons) degree in Electrical and Electronics Engineering from Birla Institute of Technology and Science, Pilani, India, in 2015 and the PhD degree from the Department of Computing, The Hong Kong Polytechnic University, Hong Kong, in 2021. He is currently an Assistant Professor of Electrical Engineering at the College of Engineering and Computer Science, VinUniversity, Hanoi, Vietnam. Prior to that, he worked at The Hong Kong Polytechnic University as a Research Fellow from July 2025 to December 2025, and as a Research Assistant Professor from April 2022 to June 2025. His research interests include Edge Computing and Intelligence, Internet of Things, and Smart Buildings. He is currently a member of IEEE and ACM.
\end{IEEEbiographynophoto}

\begin{IEEEbiographynophoto}{Jiannong Cao} (Fellow, IEEE) (jiannong.cao@polyu.edu.hk) is currently the Vice President (Education), Otto Poon Charitable Foundation Professor in Data Science, and the Chair Professor of Distributed and Mobile Computing in the Department of Computing at The Hong Kong Polytechnic University (PolyU), Hong Kong. His research interests include distributed systems and blockchain, wireless sensing and networking, big data and machine learning, and mobile cloud and edge computing. He has published 5 co-authored and 9 co-edited books, and over 800 papers in major international journals and conference proceedings. He is currently a member of Academia Europaea, a fellow of the Hong Kong Academy of Engineering Science, a fellow of IEEE, a fellow of China Computer Federation (CCF), and an ACM distinguished member. 
\end{IEEEbiographynophoto}

\begin{IEEEbiographynophoto}{Fu Xiao} (linda.xiao@polyu.edu.hk) received her double Bachelor degrees in Heating, Ventilation and Air-Conditioning Engineering (major) and Marketing (minor) from Xi'an Jiao Tong University in 1998, her Master degree in Refrigeration and Cryogenics Engineering from Shanghai Jiao Tong University in 2001, and her PhD in Building Services Engineering from the Hong Kong Polytechnic University (PolyU) in 2004. She joined PolyU as a lecturer in July 2006 and was promoted to Assistant Professor in September 2009, Associate Professor in July 2013, and Professor in July 2020. Her research interests include smart buildings and smart energy system management, with the focus on dynamic modeling of building energy systems, optimal control and diagnosis of building and urban energy systems, as well as big data analytics and AI for smart energy-efficient and resilient buildings. 
\end{IEEEbiographynophoto}

\vfill

\end{document}